\documentstyle[aps,epsf,eqsecnum]{revtex}
\begin{document}
\draft

\thispagestyle{empty}
{\baselineskip0pt
\leftline{\large\baselineskip16pt\sl\vbox to0pt{\hbox{\it Department of Physics}
               \hbox{\it Kyoto University}\vss}}
\rightline{\large\baselineskip16pt\rm\vbox to20pt{\hbox{KUNS 1512}
               \hbox{June 1999}
\vss}}%
}

\begin{center}{\large \bf 
Gravitational Radiation from a Naked Singularity}
\end{center}
\begin{center}{{\it --- Odd-Parity Perturbation ---}}
\end{center}

\begin{center}
 {\large 
Hideo Iguchi{\footnote{e-mail: iguchi@tap.scphys.kyoto-u.ac.jp}}, 
Tomohiro Harada{\footnote{e-mail: harada@tap.scphys.kyoto-u.ac.jp}},
and Ken-ichi Nakao{\footnote{e-mail: nakao@tap.scphys.kyoto-u.ac.jp}}
} \\
{\em Department of Physics,~Kyoto University, Kyoto 606-8502,~Japan} \\

\end{center}

\begin{abstract}
It has been suggested that a naked singularity
is a good candidate for a strong gravitational wave burster.
A naked singularity occurs in the generic collapse
of an inhomogeneous dust ball.
We study the odd-parity mode of gravitational waves from a naked
singularity of the Lema\^{\i}tre-Tolman-Bondi spacetime. 
The wave equation for gravitational waves is solved 
by numerical integration using the single null coordinate.
It is found that the naked singularity is not a strong source 
of the odd-parity gravitational radiation, although
the metric perturbation grows in the central region.
Therefore, the Cauchy horizon in this spacetime should be marginally
stable with respect to odd-parity perturbations.
\end{abstract}

\section{Introduction}
\label{sec:intro}
The study involved in the attempt to detect gravitational waves represents
one of the most
exciting field in modern physics. The detection will provide a
decisive test of general relativity and open a new window for
astrophysical observation. Unfortunately, in spite
a great deal of effort, no one has yet succeeded in the detection
of the gravitational waves. 
However ground-based laser interferometers, such as
LIGO,\cite{abramovici1992} VIRGO,\cite{bradaschia1990}  
TAMA\cite{kuroda1997} and GEO600,\cite{hough1992} are being
constructed and 
will soon enter a stage of data taking.
Strong sources of gravitational waves should be regions where 
gravity is general relativistic and where the velocities of bulk motion
are near  the speed of light. Many researchers have theoretically
investigated coalescing and colliding black holes and neutron stars as
candidates for detection. In addition, gravitational waves
emitted during stellar collapse to black holes have been
studied by perturbative calculations of a spherically symmetric 
background.\cite{CPM,Seidel} \ Because of the existence of event horizon,
the emitted 
gravitational waves are dominated by the quasi-normal modes of the
Schwarzschild black hole in the asymptotic region. 
Some other realistic candidates exist, e.g., supernovae, rapidly
rotating neutron stars, relic gravitational waves of the early universe,
stochastic gravitational waves from cosmic strings, and so on.

In this paper we attempt to investigate whether a naked singularity, if such
exists, is a strong source of gravitational radiation, 
and, moreover, to understand the dynamics and 
observational meaning of naked singularity formation.
Several researchers have shown that the final fate of gravitational 
collapse is not always a singularity covered by an event horizon. 
For example, in the Lema\^{\i}tre-Tolman-Bondi (LTB) 
spacetime,\cite{Tolman34,Bondi}  a
naked shell-focusing singularity appears from generic initial data for
spherically symmetric configurations of the rest mass density and a
specific energy of the dust
fluid.\cite{Eardley,Christodoulou,Newman,Joshi-Dwivedi} \
The initial functions in the most general expandable form have been
considered.\cite{Jhingan:1997ia} \ 
The matter content in such a
spacetime may satisfy even the dominant energy condition.
In this case with a small disturbance of the spacetime, very short
wavelength gravitational waves, which are  
created in the high density region around a singularity,
may propagate to the observer outside the dust cloud because of the
absence of an event horizon. If this is true,  
extremely high energy phenomena which cannot be realized by
any high energy experiment on Earth can be observed. 
In this case information regarding the physics of so-called `quantum gravity'
may be obtained. Also, these waves may be
so intense that they destroy the Cauchy horizon.
In this paper the generation of gravitational waves during the collapse of
spherical dust ball with small rotational motion is considered.

Nakamura, Shibata and Nakao\cite{nsn1993} have suggested that a 
naked singularity may emit considerable gravitational wave radiation.
This was proposed from
the estimate of gravitational radiation from a
spindle-like naked singularity.
They modeled the spindle-like 
naked singularity formation in gravitational collapse
by the Newtonian prolate dust collapse for a sequence of 
general relativistic, momentarily static initial data. It should be noted
that the system they
considered is different from that considered in this
article and that their result is controversial. There
are numerical analyses that both support and do not support the results of 
Nakamura, Shibata and
Nakao for prolate collapse\cite{Shapiro} and for cylindrical 
collapse.\cite{Echeverria:1993wf,Chiba}

Due to the non-linear nature of the problem, 
it is difficult to analytically solve the Einstein equation. Therefore, 
numerical methods will provide the final tool. However, 
its singular behavior makes 
accurate numerical analysis very difficult at some stage. 
In this article, we investigate odd-parity linear 
gravitational waves from the collapse of an inhomogeneous
spherically symmetric dust cloud. 
 Even for the linearized Einstein equation, we must perform 
the numerical integration. 
However, in contrast to the numerical simulation of 
the full Einstein equation, 
high precision is guaranteed for the numerical integration of
the linearized Einstein
equation, even in the region with extremely large spacetime curvature. 
Furthermore, the linear stability of known examples 
of naked singularity formation is necessary 
as a first step to understand the general dynamics near the ``naked 
singularity formation.''

Recently, Iguchi, Nakao and Harada 
\cite{Iguchi} (INH) studied odd-parity metric
perturbations around a naked singularity in the LTB spacetime. In INH,
it was found that the propagation of odd-parity gravitational waves is
not affected by collapse of a dust cloud, even if there appears a
central naked singularity.  When we consider the generation of 
gravitational waves from the dust collapse, we should analyze the
perturbations including their matter part. Here we investigate the
evolution equation of the odd-parity quadrupole mode for metric and matter
perturbations. This matter perturbation relates to small rotational
motion, and it produces gravitational waves during the
collapse. 
We follow the derivation of the evolution equations of the perturbations
and the numerical method to integrate these equations of INH. We
investigate the time evolutions of the gauge invariant metric variables at
the symmetric center, where a naked singularity appears, and at a constant
circumferential radius $R$. We show that the gauge
invariant variable diverges only at the center and it does not propagate to 
the outside.

As we know, the LTB spacetime is one candidate as a  
counterexample of the cosmic censorship hypothesis (CCH), which was
introduced by Penrose.\cite{Penrose69,Penrose79} \
The CCH is a very helpful assumption, because 
various theorems on the properties of black holes
were proved under this assumption.
Here the precise formulation and validity of cosmic censorship
is not our concern. We only consider the situation 
in which the extremely high-density and large-curvature region 
can be seen by some observer from the gravitational collapse.
Such a situation can be regarded as a naked singularity in a practical 
sense since we are not yet able to predict  
phenomena beyond the Planck scale. 
However, here it should be noted that even from this practical 
point of view, the stability of the Cauchy horizon with respect to 
gravitational-wave perturbations still has an important physical
meaning. Since in the system considered here, 
the mode propagating with the speed of light is the gravitational waves 
only, the instability of the Cauchy horizon implies that the gravitational 
wave introduces an extremely large spacetime curvature along the 
null hypersurface near the Cauchy horizon from the region near the central 
singularity. Hence, if the Cauchy horizon in this system is unstable, 
naked singularity formation of this type might be a strong source of 
gravitational radiation. 

Here we comment on the problem motivated in INH, i.e., 
 ``nakedness of the naked singularity.''  
Odd-parity matter perturbations are produced by rotational motion of
the dust cloud. Therefore the growth of the matter perturbation would
cause the centrifugal force to dominate the radial motion of the
cloud. Hence, an angular momentum bounce would occur. If this occurs at the
center, the central singularity will disappear. This situation seems to be
inevitable for the dipole mode of odd-parity matter perturbations. 
 In the case of a spherically symmetric system composed of
counter-rotating particles, the angular-momentum bounce can actually prevent
the naked singularity formation.\cite{Harada} \ 
However, for a system without spherical symmetry, it is still an open 
question how the non-linear asphericity works in the final stage of 
singularity formation with rotational matter perturbations because of 
the difficulty of both the analytic and numerical treatment.
Further, if the initial matter perturbation is
sufficiently small, it might still be possible that the radius of the
spacetime curvature becomes the Planck length. 
 In this case, the behavior of other modes of perturbations 
is crucial to understand the classical dynamics
in the region of the Planck scale. 
As will be shown below, since there is no gravitational wave of the dipole 
mode and the quadrupole mode generated in the dust cloud does not
propagate to the outside, the Cauchy horizon is marginally stable 
against odd-parity perturbations originating in the aspherical 
rotational motion of matter. 

The paper is organized as follows: In \S \ref{sec:be}, the basic
equations are developed; in \S \ref{sec:results}, the numerical
results are presented; these results are discussed in \S
\ref{sec:discussion}; in \S \ref{sec:summary}, we summarize our
results. We adopt the geometrized units, in which $c=G=1$. 
The signature of the metric tensor and 
sign convention of the Riemann tensor follow 
Ref.~\cite{MTW}.

\section{Basic equations}
\label{sec:be}
We consider the evolution of odd-parity perturbations of the LTB
spacetime up to linear order. The background spacetime and
perturbation method are described in INH.
\cite{Iguchi} \ In this section we briefly review them.

Using the synchronous coordinate system, the 
line element of the background LTB spacetime can be expressed in the form  
\begin{equation}
  \label{bgmetric}
  d{\bar s}^{2}= {\bar g}_{\mu\nu}dx^{\mu}dx^{\nu}\equiv
  -dt^2+A^{2}(t,r)dr^{2}
  +R^2(t,r)(d\theta^{2}+\sin^{2}\theta d\phi^{2}).
\end{equation}
The energy-momentum tensor for the dust fluid is 
\begin{equation}
  \label{bgmatter}
  {\bar T}^{\mu\nu} = {\bar \rho}(t,r){\bar u}^{\mu}{\bar u}^{\nu},
\end{equation}
where ${\bar\rho}(t,r)$ is the rest mass density and 
${\bar u}^{\mu}$ is the 4-velocity of the dust fluid. 
In the synchronous coordinate system, the unit vector field normal to 
the spacelike hypersurfaces is a geodesic, and  
there is a freedom of which timelike geodesic field 
is adopted as the hypersurface unit normal. 
Using this freedom, we can always set ${\bar u}^{\mu}=\delta^{\mu}_{0}$, 
since the 4-velocity of the spherically symmetric 
dust fluid is tangent to an irrotational 
timelike geodesic field. 

Then the Einstein equations and the equation of motion for the 
dust fluid reduce to the simple equations 
\begin{eqnarray}
  A &=&  \frac{\partial_{r}R}{\sqrt{1+f(r)}}, \label{eq:A} \\
  {\bar \rho}(t,r) &=& \frac{1}{8\pi}
  \frac{1}{R^2 \partial_{r}R}{dF(r)\over dr},\label{eq:einstein} \\
  (\partial_{t}R)^2-\frac{F(r)}{R} &=& f(r),\label{eq:energyeq} 
\end{eqnarray}
where $f(r)$ and $F(r)$ are arbitrary functions of the radial 
coordinate, $r$. From Eq.\ (\ref{eq:einstein}), $F(r)$ is related 
to the Misner-Sharp mass function,\cite{Misner} $m(r)$, 
of the dust cloud in the manner 
\begin{equation}
  \label{mass}
  m(r) = 4\pi \int_0^{R(t,r)}{\bar \rho}(t,r)R^2dR = 4\pi
  \int_0^r{\bar \rho}(t,r)R^2\partial_{r}R dr =\frac{F(r)}{2}. 
\end{equation}
Hence Eq.\ (\ref{eq:energyeq}) might be 
regarded as the energy equation per unit mass. 
This means that  
the other arbitrary function, $f(r)$, is recognized as 
the specific energy of the dust fluid. 
The motion of the dust cloud is completely specified 
by the function $F(r)$ 
(or equivalently, the initial distribution of 
the rest mass density, ${\bar \rho}$) and the specific energy, $f(r)$. 
When we restrict our calculation to the case that the symmetric center, 
$r=0$, is initially regular, the central shell focusing singularity 
is naked if and only if 
$\partial_{r}^{2}{\bar\rho}|_{r=0}<0$ is initially 
satisfied for the marginally bound collapse, $f(r)=0$.
\cite{Singh-Joshi,Jhingan} \  
For a collapse that is not marginally bound, there exists a similar
condition as an inequality for a value depending on the functional forms 
of $F(r)$ and $f(r)$.\cite{Newman,Singh-Joshi,Jhingan} \

Let us now derive the perturbation equations.
The perturbed metric tensor is expressed in the form 
\begin{equation}
  \label{metric}
  g_{\mu\nu} = {\bar g}_{\mu\nu}+h_{\mu\nu},
\end{equation}
where ${\bar g}_{\mu\nu}$ is a metric tensor of a
spherically symmetric background spacetime 
and $h_{\mu\nu}$ is a perturbation. 
The energy-momentum tensor is written in the form  
\begin{equation}
  \label{ene-mom}
  T_{\mu\nu} = \bar{T}_{\mu\nu} + \delta T_{\mu\nu},
\end{equation} 
where $\bar{T}_{\mu\nu}$ is a background quantity and 
$\delta T_{\mu\nu}$ is a perturbation. 
By virtue of the spherical symmetry of the background spacetime, 
${\bar T}_{\mu\nu}$ can be expressed in the form 
\begin{eqnarray} 
  \label{GS-matter} 
  \bar{T}_{\mu\nu}dx^{\mu}dx^{\nu} &=& 
  \bar{T}_{ab}dx^adx^b+\frac{1}{2}\bar{T}_B^{~B}R^2(t,r)d\Omega^2, 
\end{eqnarray} 
where the subscripts and superscripts $a,b,\cdots$ represent $t$ and $r$,  
while $A,B,\cdots$ represent $\theta$ and $\phi$. 
The odd-parity perturbations of  
$h_{\mu\nu}$ and $\delta T_{\mu\nu}$ are expressed 
in the form 
\begin{eqnarray} 
  \label{GS-pmetric} 
  h_{\mu\nu} &=& \left(\begin{array}{ccc} 
                        0 & 0 & h_0(t,r)\Phi^m_{l~B} \\ 
                          & 0 & h_1(t,r)\Phi^m_{l~B} \\ 
                     \mbox{sym} &   & h_2(t,r)\chi^m_{l~AB} 
                      \end{array}  \right), 
\end{eqnarray} 
\begin{eqnarray} 
  \label{GS-pmatter} 
  \delta T_{\mu\nu} &=& \left(\begin{array}{ccc} 
                        0 & 0 & t_0(t,r)\Phi^m_{l~B} \\ 
                          & 0 & t_1(t,r)\Phi^m_{l~B} \\ 
                     \mbox{sym} &   & t_2(t,r)\chi^m_{l~AB} 
                      \end{array}  \right) , 
\end{eqnarray}  
where $\Phi^m_{l~B}$ and $\chi^m_{l~AB}$ are odd-parity vector and 
tensor harmonics associated with the spherical symmetry of 
the background spacetime.\cite{Regge} \ We set all the arbitrary
constants in the definitions of harmonics to unity.
We then introduce the gauge-invariant 
variables defined by Gerlach and Sengupta.\cite{Gerlach-Sengupta} \ 
The metric variables are given by 
\begin{equation} 
  \label{ka} 
  k_a = h_a-\frac{1}{2}R^2\partial_a\left(\frac{h_2}{R^2}\right).
\end{equation} 
The matter variables are given by the combinations 
\begin{eqnarray}
  \label{la}
  L_a &=& t_a-\frac{1}{2}\bar{T}_B^{~B}h_a ,\\
  \label{l}
  L &=& t_2-\frac{1}{2}\bar{T}_B^{~B}h_2 .
\end{eqnarray}
In the LTB case, the odd-parity gauge-invariant
matter variables become 
\begin{eqnarray}
  \label{La-L}
      L_0 = \bar{\rho}(t,r) U(t,r)~~~~~~{\rm and}~~~~~~ L_1 = L = 0,
\end{eqnarray}
where $U(t,r)$ represents the perturbation of the 4-velocity as
$\delta u_\mu = (0,0,U(t,r)\Phi_{l~B}^m)$.
The evolution equation for the matter variable (Eq.\ (3$\cdot$19) in INH),
\begin{equation}
  \label{T-Bconserve}
  \partial_t\left(AR^2L_0\right) = 0,
\end{equation}
is easily integrated, and we obtain
\begin{equation}
  \label{kakuundouryou}
  L_0 = \frac{1}{AR^2}{dJ(r)\over dr},
\end{equation}
where $J(r)$ is an arbitrary function depending only on 
$r$. From Eqs.\ (\ref{eq:einstein}), (\ref{La-L}), and (\ref{kakuundouryou}),
we obtain the relation
\begin{eqnarray}
  \label{U(r)}
  U(t,r)&=&8\pi \sqrt{1+f(r)}\frac{dJ(r)/dr}{dF(r)/dr} \\
        &\equiv& U(r),
\end{eqnarray}
so $U(t,r)$ is independent of the time coordinate $t$.
We introduce a gauge-invariant variable for the metric as
\begin{equation}
  \label{psis}
  \psi_s \equiv {1\over A}\left[\partial_t
  \left(\frac{k_1}{R^2}\right)-\partial_r
  \left(\frac{k_0}{R^2}\right)\right].
\end{equation}
The metric perturbation variables, $k_0$ and $k_1$, are reconstructed
from the linearized Einstein equations,
\begin{eqnarray}
  \label{wave2}
  \partial_r\left(R^4\psi_s\right)+A\left(l-1\right) 
\left(l+2\right)k_0 &=& 16\pi AR^2L_0, \\
  \label{wave3}
  \partial_t\left(R^4\psi_s\right)+\frac{1}{A}
\left(l-1\right)\left(l+2\right)k_1 &=& 0,  
\end{eqnarray}
by substituting $\psi_s$. It was
shown in INH that $\psi_s$ is closely connected to the tetrad components 
of the magnetic part of the Weyl tensor. These components diverge where
and only where $\psi_s$ diverges. From the linearized Einstein equations 
we obtain the linearized evolution equation for the odd-parity
perturbation as
\begin{eqnarray}
  \label{wave4}
  \partial_{t}\left(\frac{A}{R^2}\partial_{t}
\left(R^4\psi_s
\right)\right)&-&\partial_{r}
\left(\frac{1}{AR^2}\partial_{r}
\left(R^4\psi_s\right)\right) +
\left(l-1\right)
\left(l+2\right)A\psi_s \nonumber\\ 
&=& -16\pi\partial_{r}\left(\frac{1}{AR^2}{dJ\over dr}\right).
\end{eqnarray}

The regularity conditions at the center are also considered in INH.
The result for the matter perturbation $L_0$ is given by 
\begin{equation}
  \label{r-0-L}
  L_0 \longrightarrow L_c(t)r^{l+1}+O(r^{l+3}).
\end{equation}
Therefore the matter perturbation variables vanish at the regular center 
independent of the value of $l$.
The metric variable $\psi_s$ behaves near the center as 
\begin{eqnarray}
  \label{r-0-psis}
  \psi_{s} &\longrightarrow& 
 \psi_{sc}(t)r^{l-2}+O(r^l)
 ~~~~~~~~~~~~\mbox{for $l\geq 2$}, \\
  \label{r-0-psis1}
  \psi_{s} &\longrightarrow& \psi_{sc}(t)r+O(r^3)
  ~~~~~~~~~~~~~~~~\mbox{for $l=1$}.
\end{eqnarray}
 From the above equations, it is shown that only the quadrupole mode, 
$l=2$, of $\psi_{s}$ does not vanish at the regular center. 

\section{Results}
\label{sec:results}
We numerically solve the wave equation (\ref{wave4}) in the case of 
marginally bound collapse, $f(r)=0$, and the quadrupole mode, $l=2$. 
We follow numerical techniques employed in INH to integrate the time
evolution of the
perturbations. The numerical code is essentially the same as that in INH,
which is tested by comparison with the analytic solution for the Minkowski 
background.

By virtue of the relation $f(r)=0$, 
we can easily integrate Eq.\ (\ref{eq:energyeq}) and obtain 
\begin{equation}
  \label{f=0}
  R(t,r) = \left(\frac{9F}{4}\right)^{1/3}[t_0(r)-t]^{2/3},
\end{equation}
where $t_0(r)$ is an arbitrary function of 
$r$. The naked singularity formation time is $t_0=t_0(0)$. Using the freedom
for the scaling of $r$, we choose $R(0,r)=r$. This scaling of $r$ 
corresponds to the following choice of $t_{0}(r)$: 
\begin{equation}
  \label{t0}
  t_0(r) = \frac{2}{3\sqrt{F}}r^{3/2}.
\end{equation}
Here note that, from Eq.\ (\ref{eq:A}), the background 
metric variable, $A$, is equal to $\partial_{r}R$.  
Then, the wave equation (\ref{wave4}) becomes 
\begin{eqnarray}
  \label{wave-eq}
  \frac{\partial^2 \psi_s}{\partial t^2}-\frac{1}{(\partial_{r}R)^2}
 \frac{\partial^2 \psi_ s}{\partial  r^2}
  &=& \frac{1}{(\partial_{r}R)^{2}}\left(6\frac{\partial_{r}R}{R}
  -\frac{\partial_{r}^{2}R}{\partial_{r}R}\right)\frac{\partial
  \psi_s}{\partial r}-\left(6\frac{\partial_{t}R}{R}
  +\frac{\partial_{t}\partial_{r}R}{\partial_{r}R}\right)
  \frac{\partial
  \psi_s}{\partial t} \nonumber \\ 
& &-4\left[\left(\frac{\partial_{t}\partial_{r}R}{\partial_{r}R}\right)
  \frac{\partial_{t}R}{R}
  +\frac{1}{2}\left(\frac{\partial_{t}R}{R}\right)^2\right]\psi_s  \nonumber \\
& &-\frac{16\pi}{(\partial_r R)R^2}\partial_{r}
\left(\frac{r^2 \rho (r) U(r)}{(\partial_r R)R^2}\right),
\end{eqnarray}
where $\rho (r) = \bar{\rho}(0,r)$ is the density profile at $t=0$.
We solve this partial differential equation numerically. 

Before getting into the detailed explanation of the numerical techniques, we
comment on the behavior of the matter perturbation variable $L_0$ around 
a naked singularity on the slice $t=t_0$.
The regularity conditions of $L_0$ and $\bar{\rho}$
determine the behavior of $U(r)$ near the center as
\begin{equation}
  \label{dJdr-c}
  U(r) \propto r^{l+1}.
\end{equation}
This property does not change even if a central singularity
appears. However, the $r$ dependence of $R$ and $A$ near the center changes at 
that time. Assuming a rest mass density profile of the form
\begin{equation}
  \rho(r) = \rho_0 + \rho_n r^n + \cdots ,
\end{equation}
we obtain the relation 
\begin{equation}
  t_0(r) \propto t_0 + t_n r^n
\end{equation}
 from Eqs.\ (\ref{eq:einstein}) and (\ref{t0}),
where $n$ is a positive even integer. After substituting this 
relation into Eq.\ (\ref{f=0}),
the lowest order term is absent from the square brackets of it.
Then we obtain the behavior of $R$ and $A$ around the central singularity as  
\begin{equation}
  \label{singr}
  R(t_0,r) \propto r^{1+ {2\over3} n},
\end{equation}
and
\begin{equation}
  \label{singa}
  A(t_0,r) \propto r^{{2\over3} n}
\end{equation}
on the slice $t=t_0$.
As a result, we obtain the $r$
dependence of $L_0$ around the center when the naked singularity appears as
\begin{equation}
  L_0(t_0,r) \propto r^{l-2n+1}.
\end{equation}
For example, if $l=2$ and $n=2$, then $L_0$ is inversely proportional to $r$
and diverges toward the central naked singularity. Therefore the source term
of the wave equation is expected to have a large magnitude around the naked 
singularity. Thus the metric perturbation variable $\psi_s$ as well as
matter variable $L_0$ may diverge toward the naked singularity.

Instead of the $(t,r)$ coordinate system, we 
introduce a single-null coordinate system, $(u,r')$,
where $u$ is an outgoing null coordinate and chosen so that it
agrees with $t$ at the symmetric center and we choose $r'=r$.
We perform the numerical integration along two characteristic
directions. Therefore we use a double null grid in the numerical
calculation. A detailed explanation of these coordinates is given in
INH. By using this new coordinate system, $(u,r')$, Eq.\ (\ref{wave-eq}) 
is expressed in the form 
\begin{eqnarray}
  \label{dphis/dlambda}
  \frac{d\phi_s}{du} &=&
  -\frac{\alpha}{R}\left[3\partial_{r}R
  +{1\over2}R(\partial_{t}R)\partial_{t}\partial_{r}R
  -\frac{5}{4}(\partial_{t}R)^{2}\partial_{r}R\right]\psi_s \nonumber \\
    & &
  -\frac{\alpha}{2}\left[\frac{\partial_{r}^{2}R}{(\partial_{r}R)^2}
  -\frac{2}{R}\left(1-\partial_{t}R\right)
   \right]\phi_s -\frac{8\pi \alpha}{R}\partial_{r}\left(
   \frac{r^2 \rho (r) U(r)}{(\partial_rR)R^2}\right),\\
  \label{delpsi/delrprime}
  \partial_{r'}\psi_{s}&=& \frac{1}{R}\phi_s
  -3\frac{\partial_{r}R}{R}
\left(1+\partial_{t}R\right)\psi_s,
\end{eqnarray}
where the ordinary derivative on the left-hand side of
Eq.\ (\ref{dphis/dlambda}) and the partial derivative on the left-hand
side of Eq.\ (\ref{delpsi/delrprime}) are given by
\begin{eqnarray}
  \frac{d}{du} &=& \partial_u +
  \frac{dr'}{du}\partial_{r'}
   = \partial_u-\frac{\alpha}{2\partial_r R}\partial_{r'} \nonumber  \\ 
  &=&\frac{\alpha}{2}\partial_t-\frac{\alpha}{2\partial_r R}\partial_{r},\\
  \label{r'a}
  \partial_{r'} &=& -\frac{(\partial_r u)_t}{(\partial_t
    u)_r}\partial_{t}+\partial_{r}=
    (\partial_{r}R)\partial_{t}+\partial_{r},
\end{eqnarray}
respectively. Also, $\phi_s$ is defined by 
Eq.\ (\ref{delpsi/delrprime}) and $\alpha$ is given by
\begin{equation} 
  \alpha\equiv {1\over (\partial_{t} u)_r}.
\end{equation}
We integrate Eq.\ (\ref{dphis/dlambda}) using the scheme of an explicit first
order difference equation. We use the trapezoidal rule,  
\begin{equation}
  \psi_{s j+1}=\psi_{s j}+\frac{\Delta r'}{2}\left((\partial_{r'}\psi_{s})_{j}+(\partial_{r'}\psi_{s})_{j+1}\right),
\end{equation}
to integrate Eq.\ (\ref{delpsi/delrprime}).

For the boundary condition at the center we demand that $\psi_s$ behaves
as $\psi_{sc}(t)+\psi_{s2}(t)r^2$ on a surface of $t=\mbox{const}$. We
numerically realize this condition by two-step interpolation. First
the values of $\psi_s$ are derived at two points on the surface of
$t=\mbox{const}$ from the interpolation on the slices of $u=\mbox{const}$. 
Next, using 
these two values, the central value of $\psi_s$ is derived from the
interpolation on the slice of $t=\mbox{const}$. 
Another way to determine the central value of 
$\psi_s$ is as follows. We first obtain the central value of $\phi_s$
from Eq.\ (\ref{dphis/dlambda}). From Eq.\ (\ref{delpsi/delrprime}) and
the boundary conditions, the relation of $\psi_s$ and $\phi_s$ at the
center is given by 
\begin{equation}
 \label{phi-psi}
  \phi_s = 3 \partial_{r'}R \psi_s.
\end{equation}
Using this relation the central value of $\psi_s$ is obtained. In our
numerical analyses the results of these two methods agree well.

We assume $\psi_s$ vanishes on the initial null hypersurface. Therefore,
there
exist initial ingoing waves which offset the waves produced by the
source term on the initial null hypersurface. In INH, it is confirmed
that this type of the initial ingoing waves propagate through the dust 
cloud without net
amplification even when they pass through the cloud just before the
appearance of the naked singularity. Therefore those parts of the metric
perturbations would not diverge at the center.

We adopt the initial rest mass density profile 
\begin{equation}
  \label{density}
  \rho(r)=\rho_0 \frac{1+ \exp\left(-\frac{1}{2}\frac{r_1}{r_2}\right)}
  {1+ \exp\left(\frac{r^n -r_1^n}{2 r_1^{n-1}r_2}\right)},
\end{equation}
where $\rho_0$, $r_1$ and $r_2$ are positive constants and $n$ is a 
positive even
integer. As a result the dust fluid spreads all over the space. However,
if $r \gg r_1,r_2$, then $\rho(r)$ decreases exponentially, so that the dust
cloud is divided into the core part and the envelope which would be
considered as the vacuum region essentially. We define a core radius as
\begin{equation}
  \label{cradious}
  r_{\mbox{\scriptsize core}}=r_1+\frac{r_2}{2}.
\end{equation}
If we set $n=2$, there appears a central naked
singularity. This singularity becomes locally or globally naked depending 
on the parameters ($\rho_0,r_1,r_2$). However, if the integer
$n$ is greater than $2$, the final state of the dust cloud is a black
hole independently of the parameters. Then we consider three different
density profiles connected with three types of the final state of the
dust cloud, globally and locally naked singularities and a black
hole. The outgoing null coordinate $u$ is chosen so that it agrees
with the proper time at the symmetric center. Therefore, even if the
black hole background is considered, we can analyze the inside of the 
event horizon.
Corresponding parameters are given in Table \ref{tab:parameter}.
Using this density profile, we numerically calculate the total 
gravitational mass of 
the dust cloud $M$. In our calculation we adopt the total
mass $M$ as the unit of the variables.

The source term of Eq.\ (\ref{dphis/dlambda}),
\begin{equation}
  \label{souce}
  S(t,r)=-\frac{8\pi \alpha}{R}\partial_{r}\left(
   \frac{r^2 \rho (r) U(r)}{(\partial_rR)R^2}\right),
\end{equation}
is determined by $U(r)$. As mentioned above, the constraints on the 
functional form of
$U(r)$ are given by the regularity condition of $L_0$. From
Eq.\ (\ref{dJdr-c}), $U(r)$ should be
proportional to $r^{l+1}$ toward the center. 
We localize the matter perturbation near the center to diminish the
effects of the initial ingoing waves. Therefore we define $U(r)$ such that
\begin{equation}
  \label{dJdr}
  r^2 \rho(r) U(r) = \left\{ 
     \begin{array}{ccc}
      U_0 \left( \frac{r}{r_b} \right)^5 \left(1-\left
  ( \frac{r}{r_b} \right)^2 \right)^5  &{\mbox{for}} & 0\leq r\leq r_b,\\
        0 &{\mbox{for}} & r>r_b,
      \end{array}
    \right.
\end{equation}
where $U_0$ and $r_b$ are arbitrary constants. In our numerical calculation
we chose $r_b$ as $r_{\mbox{\scriptsize core}}/2$. This choice of $r_b$ has no
special meaning, and the results of our numerical calculations are not
sensitive to it.

\begin{figure}
  \begin{center}
    \leavevmode
    \epsfysize=250pt\epsfbox{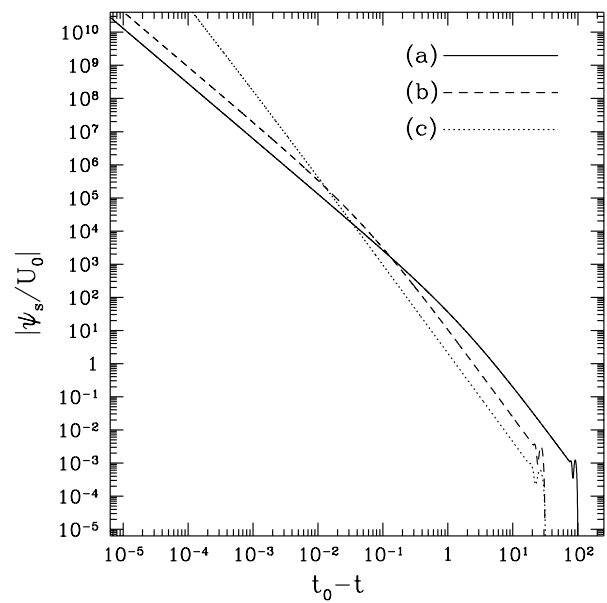}    
 \caption{Plots of $\psi_s$ at the center as a function of the time coordinate
   $t$. The solid line represents the globally naked case (a), the dashed
   line represents the locally naked case (b), and the dotted line
   represents the black hole case (c).}
  \label{fig:center}
  \end{center}
 \end{figure}

\begin{figure}
  \begin{center}
    \leavevmode
    \epsfysize=250pt\epsfbox{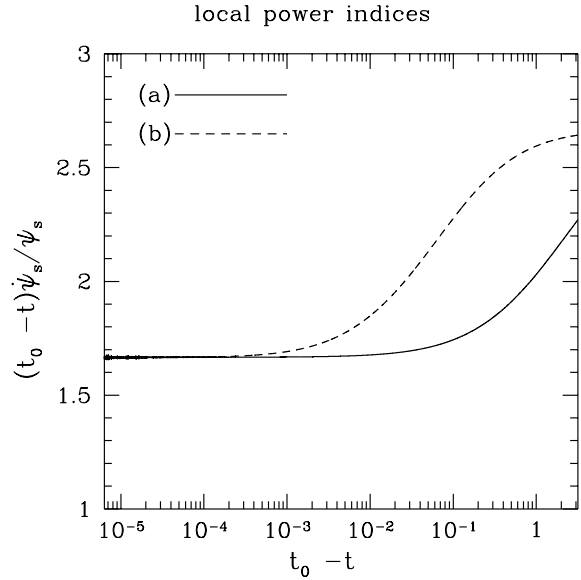}    
  \caption{Plots of the local power indices
    $(t_0-t)\dot{\psi_s}/\psi_s$. The solid line corresponds to the globally
    naked case (a), and the dashed line corresponds to the locally naked case
    (b). Both of them approach a value near 5/3.}
 \label{fig:index}
  \end{center}
 \end{figure}

First we observe the behavior of $\psi_s$ at the center. The results are 
plotted in Fig.\ \ref{fig:center}. The initial oscillations correspond to
the initial ingoing waves. After these oscillations, $\psi_s$  grows
proportional to $(t_0-t)^{- \delta}$ for the naked singularity cases
near the formation epoch of the naked singularity. 
For the case of black hole formation, $\psi_s$ exhibits power-law
growth in the early part. Later its slope gradually changes 
but it grows faster than in the case of the naked singularity. 
For the naked singularity cases, the power-law indices $\delta$ are
determined by $(t_0-t)\dot{\psi_s}/\psi_s$ locally. The results are
shown in Fig.\ \ref{fig:index}. From this figure we read the final
indices as 5/3 for both naked cases. 
Therefore the metric perturbations diverge at the central naked singularity.

 \begin{table}
 \caption{Parameters of initial density profiles, power law indices, and 
    damped oscillation frequencies.}
 \label{tab:parameter}
 \begin{tabular}{cccccccc}
  & final state & $\rho_0$ & $r_1$ & $r_2$ & $n$ &power index&
 damped oscillation frequency\\ \hline
 (a) & globally naked & $1 \times 10^{-2}$ & 0.25 & 0.5 & 2 & 5/3 & --- \\ 
 (b) & locally naked & $1 \times 10^{-1}$ & 0.25 & 0.5 & 2 & 5/3 & 0.37+0.089$i$\\ 
 (c) & black hole & $2 \times 10^{-2}$ & 2 & 0.4 & 4 & --- & 0.37+0.089$i$\\ 
 \end{tabular}
 \end{table}

\begin{figure}
  \begin{center}
    \leavevmode
    \epsfysize=250pt\epsfbox{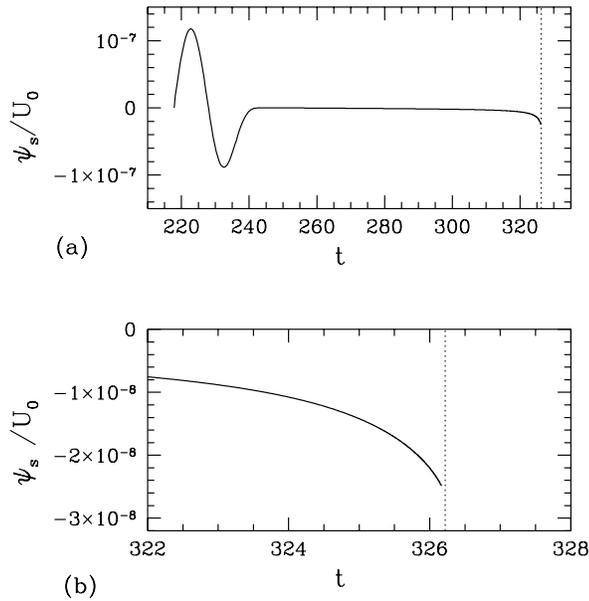}    
  \caption{Plots of $\psi_s$ for the globally naked case
    (a) at $R=100$. In (a) the left hand side
    oscillation originates from the initial ingoing wave. In (b) we
    magnify the the right-hand edge, which is just before the Cauchy 
    horizon. The dotted lines represent the time at which the observer
    at $R=100$ 
    intersects the Cauchy horizon, which is determined by numerical
    integration of the null geodesic equation from the naked singularity.}
 \label{fig:outsideg}
  \end{center}
 \end{figure}

\begin{figure}
  \begin{center}
    \leavevmode
    \epsfysize=250pt\epsfbox{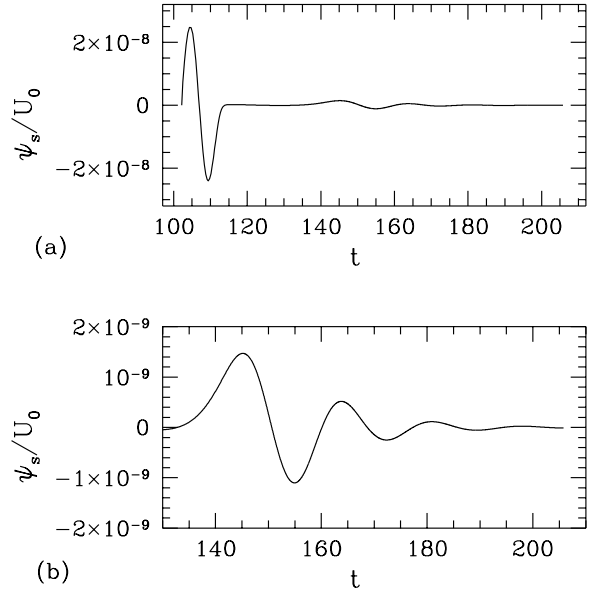}    
  \caption{Plots of $\psi_s$ for the locally naked case (b) at $R=100$. 
    In (a) the left hand side
    oscillation originates from the initial ingoing wave. After this
    oscillation, the damped oscillation dominates, and this part of
    the wave form  is magnified in (b). }
 \label{fig:outsidel}
  \end{center}
 \end{figure}

\begin{figure}
  \begin{center}
    \leavevmode
    \epsfysize=250pt\epsfbox{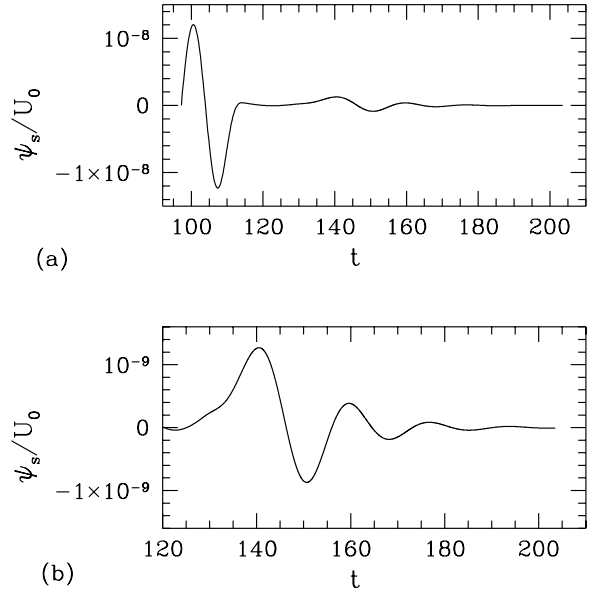}    
  \caption{Plots of $\psi_s$ for the black hole case (c) at $R=100$. 
    In (a) the left-hand side
    oscillation originates from the initial ingoing wave. After this
    oscillation, the damped oscillation dominate, as depicted
    in (b). }
 \label{fig:outsideb}
  \end{center}
 \end{figure}

We also observe the wave form of $\psi_s$ along the line of a
constant circumferential radius outside the dust cloud. The results are
shown in Figs.\ \ref{fig:outsideg}--\ref{fig:outsideb}. Figure 
\ref{fig:outsideg} displays the wave form of the globally naked case
(a), Fig.\ \ref{fig:outsidel} displays the wave form of the
locally naked case (b), and Fig.\ \ref{fig:outsideb} displays the
wave form of the black hole case (c).
The initial oscillations
correspond to the initial ingoing waves. In the case of a locally naked
singularity and black hole formation, damped oscillations dominate the 
gravitational waves. We read the frequencies and 
damping rates of these damped oscillations from Figs.\
\ref{fig:outsidel} and \ref{fig:outsideb} and give them in terms of 
complex frequencies as $0.37+0.089i$ for locally naked
and black hole cases. These agree well with the
fundamental quasi-normal frequency of the quadrupole mode
$(2M\omega = 0.74734 + 0.17792 i)$ of a Schwarzschild black hole given
by Chandrasekhar and Detweiler.\cite{Chandra} \ In the globally naked
singularity case (a), we did not see this damped oscillation  
because of the
existence of the Cauchy horizon. In all cases the gravitational waves 
generated by matter perturbations are at most quasi-normal modes of a
black hole, which is generated outside the dust cloud. 
Therefore intense odd-parity gravitational waves would not be
produced by the inhomogeneous dust cloud collapse. We should not expect
that the central extremely high density region can be observed by this
mode of gravitational waves.

We can calculate the radiated power of the gravitational waves and thereby
grasp the physical meaning of the gauge-invariant
quantities.\cite{CPM} \ To relate the perturbation of the metric to the 
radiated
gravitational power, it is useful to specialize to the radiation gauge,
in which the tetrad components $h_{(\theta)(\theta)}-h_{(\phi)(\phi)}$
and $h_{(\theta)(\phi)}$ fall off as $O(1/R)$, and all other tetrad
components fall off as $O(1/R^2)$ or faster. Note that, in vacuum at
large distance, the spherically symmetric background metric is given by
the Schwarzschild solution, where hereafter we adopt the Schwarzschild
coordinates,  
\begin{equation}
  \label{Schmetric}
  ds^2 = -\left( 1-\frac{2M}{R}\right)d\tau ^2 +\left
  ( 1-\frac{2M}{R}\right)^{-1} dR^2 + R^2 \left( 
  d\theta^{2}+\sin^{2}\theta d\phi^{2}\right) . 
\end{equation}
The relation between the line elements Eq.\ (\ref{bgmetric}) and
Eq.\ (\ref{Schmetric}) is given by the transfer matrix 
\begin{eqnarray}
  d\tau &=& \frac{1}{1-(\partial_t R)^2} (dt+\partial_r R \partial_t R
  dr), \\
  dR &=& \partial_t Rdt + \partial_r R dr.
\end{eqnarray}
In this gauge, the metric perturbations in Eq. (\ref{GS-pmetric}) behave
as 
\begin{eqnarray}
  h_0, h_1 &=& O\left(\frac{1}{R}\right), \\
  h_2 &=& w(\tau - R_*) + O(1),
\end{eqnarray}
where 
\begin{equation}
  R_* = R + 2M \ln \left( \frac{R}{2M} - 1\right) + \mbox{const}.
\end{equation}
Then, the gauge-invariant metric perturbations (\ref{ka}) are calculated 
as 
\begin{eqnarray}
  k_0 &=& -\frac{1}{2}w^{(1)}R + O(1), \\
  k_1 &=& \frac{1}{2}w^{(1)}R + O(1),
\end{eqnarray}
where $w^{(1)}$ denotes the first derivative of $w$ 
with respect to its argument.

\begin{figure}
   \begin{center}
     \leavevmode
    \epsfysize=250pt\epsfbox{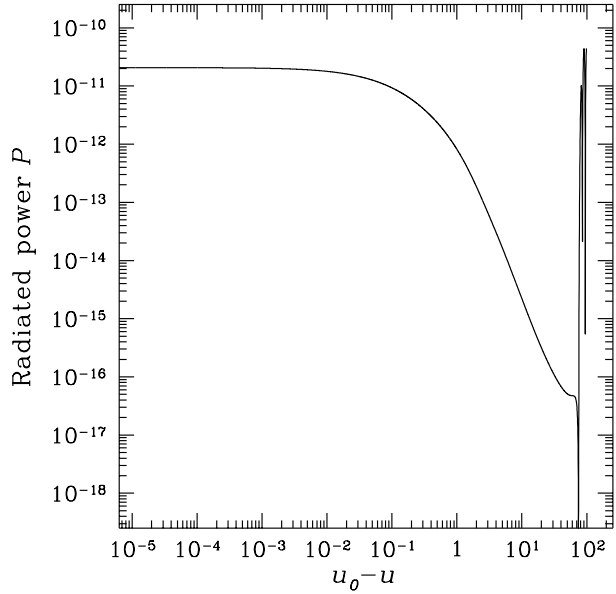}    
     \caption{Plots of the radiated power $P$ for globally naked case
       (a) at $R=100$. The horizontal axis is the out-going null
       coordinate $u$. At the Cauchy horizon, this coordinate has the
       value $u_0$.}
     \label{fig:power}
   \end{center}
 \end{figure}

In this radiation gauge, the radiated power $P$ per unit solid angle is
given by the formula which was derived by Landau and Lifshitz
\cite{Landau} from their stress-energy pseudo-tensor:
\begin{equation}
  \frac{dP}{d\Omega}=\frac{R^2}{16\pi}\left[\left(\frac{\partial
  h_{(\theta)(\phi)}}{\partial \tau}\right)^2
  +\frac{1}{4}\left(\frac{\partial h_{(\theta)(\theta)}}{\partial
  \tau}-\frac{\partial h_{(\phi)(\phi)}}{\partial \tau}\right)^2\right].
\end{equation}
For the axisymmetric mode, i.e., $m=0$, the above formula is reduced to 
\begin{equation}
  \frac{dP}{d\Omega}=\frac{1}{64\pi}(w^{(1)})^2A_l (\theta),  
\end{equation}
where
\begin{equation}
  A_l (\theta) \equiv \frac{2l+1}{4\pi}\sin^4\theta\left(\frac{d^2P_l(\cos\theta)}{(d\cos\theta)^2}\right)^2.
\end{equation}
Then, by using the gauge-invariant quantities and integrating over the
all solid angles, the formula for the power of gravitational radiation
is obtained in the following form:
\begin{eqnarray}
  \frac{dP}{d\Omega}&=&\frac{1}{16\pi}\frac{k_0^2}{R^2}A_l
  (\theta)=\frac{1}{16\pi}\frac{k_1^2}{R^2}A_l (\theta),\\
  P &=& \frac{1}{16\pi} B_l \frac{k_0^2}{R^2} = \frac{1}{16\pi} B_l
  \frac{k_1^2}{R^2} ,  
\end{eqnarray}
where
\begin{equation}
  B_l \equiv \frac{(l+2)!}{(l-2)!}.
\end{equation}
Using Eq. (\ref{wave3}), the radiated power $P$ of the quadrupole mode 
is given by
\begin{equation}
  P = \frac{3}{32\pi}R'^2\left[\partial_{\tau}\left(R^3 \psi_s\right)\right]^2.
\end{equation}
Figure \ref{fig:power} displays the time evolution of the radiated power
$P$. The radiated power also has a finite value at the
Cauchy horizon. The total energy radiated by odd-parity quadrupole
gravitational waves during the dust collapse should not diverge.

\section{Discussion}
\label{sec:discussion}

\begin{figure}
  \begin{center}
    \leavevmode
    \epsfysize=250pt\epsfbox{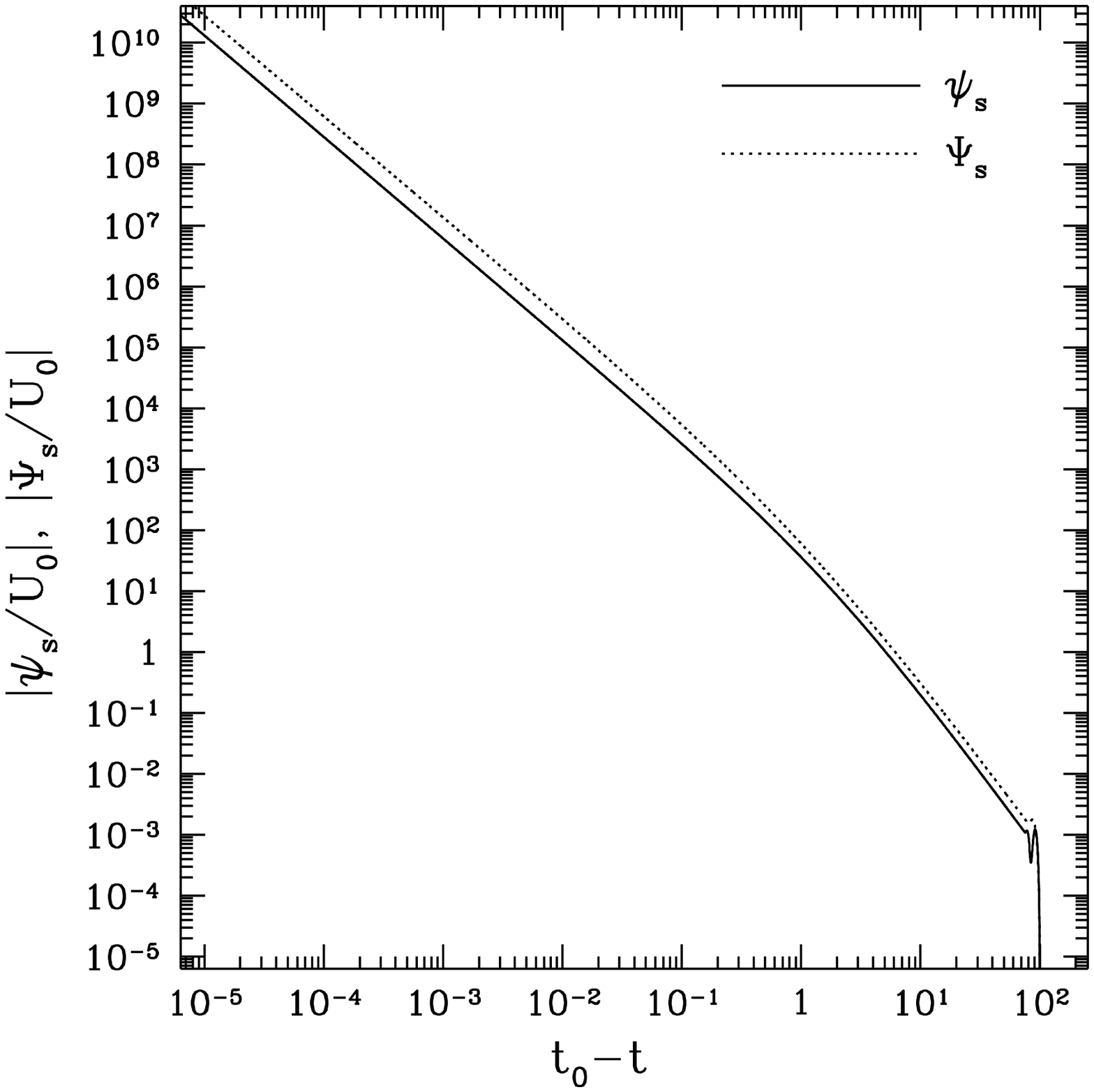}    
  \caption{Plots of $\psi_s$ and the estimated value $\Psi_s$ at the
    center for the globally naked case (a). The 
    solid line represents the $\psi_s$, and the dotted line represents the
    estimated value $\Psi_s$. Both lines exhibit power-law behavior with power
    indices 5/3. }
 \label{fig:int}
  \end{center}
 \end{figure}

First we consider the behavior of the source term $S(t,r)$ around the
naked singularity. From the regularity conditions and
Eqs.\ (\ref{dJdr-c}), (\ref{singr}), and (\ref{singa}), 
the asymptotic behavior of the source term is obtained as
\begin{equation}
  S(t,r) \propto r^{l-1}
\end{equation}
for $t<t_0$ and
\begin{equation}
  S(t,r) \propto r^{l-\frac{8}{3}n-1},
\end{equation}
at $t=t_0$. For example,
in the case $l=2$ and $n=2$, the souce term
behaves on $t=t_0$ as
\begin{equation}
  S(t,r) \propto r^{-13/3},
\end{equation}
and then it diverges at the center. Thus the divergency of $\psi_s$ at the 
center originates from the source term. To confirm this,
we numerically integrate the source term along the ingoing null lines 
with respect to $u$ and estimate the central value of $\phi_s$. We define this
`estimated' value as 
\begin{equation}
  \Phi_s \equiv \int S(t,r) du.
\end{equation}
Using Eq.\ (\ref{phi-psi}) we can define the estimated value of $\psi_s$ as 
\begin{equation}
  \Psi_s \equiv \frac{\Phi_s}{3 \partial_rR}.
\end{equation}
We plot it in Fig.\ \ref{fig:int} together with the corresponding
$\psi_s$. The estimated value has the same power-law index of $\psi_s$. 
We conclude that the behavior of $\psi_s$ is determined 
by the source term in the dust cloud.

We next consider the stability of the Cauchy horizon.
We found that the metric perturbation produced by the source term 
does not propagate outside the dust cloud, except for quasi-normal ringing. 
The source term, which controls $\psi_s$, does not diverge at the Cauchy
horizon. Therefore $\psi_s$ should not diverge at the Cauchy horizon and
should not
destroy it. Then, even if odd-parity perturbations are considered, it
will not be the case that the LTB spacetime 
loses its character as
a counterexample to
CCH due to Cauchy horizon instability. Also, it does not seem that such
collapse is a strong source of gravitational waves.

In this paper, we have dealt with the marginally bound
case. For the case of non-marginally bound collapse, 
the condition of the appearance
of the central naked singularity is slightly different from that in 
the above
case\cite{Singh-Joshi,Jhingan} and hence there is the possibility 
that the behavior of $\psi_{s}$ in this case is different from that in the 
marginally bound case. However, it is well known that 
the limiting behavior of the metric as
$t \rightarrow t_0(r)$ is common for all the cases:\cite{Landau} 
\begin{equation}
  R \approx \left(\frac{9F}{4}\right)^{1/3}\left(t_0-t\right)^{2/3}
  ,~~~A \approx \left(\frac{2F}{3}\right)^{1/3} \frac{t_0'}{\sqrt{1+f}}
  \left(t_0-t\right)^{-1/3}.
\end{equation}
We conjecture that the results of the 
perturbed analysis for the non-marginal collapse would be similar to 
the results for the marginal bound.

\section{Summary}
\label{sec:summary}
We have studied the behavior of the odd-parity perturbation
in the LTB spacetime including the matter perturbation. 
For the quadrupole mode, where gravitational waves exist, we
have numerically investigated the wave equation. 
For the case of naked singularity formation,
the gauge-invariant metric variable, $\psi_s$, diverges according to a power 
law with power index 5/3 at the center. This power index is closely
related to the behavior of the matter perturbation around the
center. We have also observed $\psi_s$ at a
constant circumferential radius. For the globally naked case, we cannot
see intense gravitational waves propagated from the center just before
the crossing of the Cauchy horizon. For the locally naked case, we
have confirmed that there exist quasi-normal oscillations. As a
result, we conclude that the type of singularity changes
due to the odd-parity perturbation because $\psi_s$ diverges at the
center. However, the Cauchy horizon is marginally stable against  
odd-parity perturbations.

At the final stage of the collapse, the effects of the rotational motion 
are important and the centrifugal force might dominate the radial motion. 
 If this is true, the central singularity would disappear when an
odd-parity matter perturbation is introduced. 
For the dipole mode, such a situation seems to be inevitable. 
 However, we should note that it is a non-trivial and open question 
how non-linear asphericity affects the final fate of the 
singularity-formation process. 
Further, in the case of initially sufficiently small aspherical
perturbations, the radius of spacetime curvature at the center might 
reach the Planck length, and hence there is still the possibility that 
the naked singularity is formed there in a practical sense.
However, as our present analysis has revealed, since the Cauchy horizon
is stable with respect to odd-parity linear perturbations,  
there is little possibility that this collapse is a strong source 
of odd-parity gravitational waves. 

There remains important related works to be completed.
The first problem is to consider the even-parity mode in which the 
metric and matter perturbations are essentially coupled with each other. 
We are now investigating this problem. 
Finally, we should consider the non-linear effects to complete this analysis. 
This problem will be analyzed in the future. 

\section*{Acknowledgements}
We would like to thank T. Nakamura for helpful and useful discussions and
N. Sugiyama for careful reading of the manuscript. We are also
grateful to H. Sato and colleagues in the theoretical astrophysics group
at Kyoto University for useful comments and encouragement.
This work was partly supported by Grants-in-Aid for Scientific
Research (No.\ 9204) and for Creative Basic Research (No.\ 09NP0801)
from the Japanese Ministry of Education,
Science, Sports and Culture.

\begin{thebibliography}{99}
 \bibitem{abramovici1992}
   A.~Abramovici et al., Science {\bf 256} (1992), 325.
 \bibitem{bradaschia1990}
   C.~Bradaschia et al., Nucl. Instrum. and Methods
   {\bf A289} (1990), 518.
 \bibitem{kuroda1997}
   K.~Kuroda et al.,
   in {\em Proceedings of International Conference on Gravitational
   Waves: Sources and Detectors}, ed. I.~Ciufolini and
   F.~Fidecaro (World Scientific, Singapore, 1997), p. 100.
 \bibitem{hough1992}
   J.~Hough,
   in {\em Proceedings of the Sixth Marcel Grossman Meeting},
   ed. H.~Sato and T.~Nakamura (World Scientific, Singapore, 1992), p. 192.
 \bibitem{CPM}
   C. T. Cunninghum, R. H. Price, and V. Moncrief, Astrophys, J. 
   {\bf 224} (1978), 643; {\bf 230} (1978), 870; {\bf 236} (1980), 674.
 \bibitem{Seidel}
   E. Seidel and T. Moore, Phys. Rev. {\bf D35} (1987), 2287. \\
   E. Seidel, E.S. Myra and T. Moore, ibid. {\bf 38} (1988), 2349. \\
   E. Seidel, ibid. {\bf 42} (1990), 1884.
 \bibitem{Tolman34}
   R. C. Tolman, Proc. Natl. Acad. Sci. U. S. A. {\bf 20} (1934), 169.
 \bibitem{Bondi}
   H. Bondi, Mon. Not. R. Astron. Soc. {\bf 107} (1947), 410.
 \bibitem{Eardley}
   D. M. Eardley and L. Smarr, Phys. Rev. {\bf D19} (1979), 2239.
 \bibitem{Christodoulou}
  D. Christodoulou, Commun. Math. Phys. {\bf 93} (1984), 171. 
 \bibitem{Newman}
   R. P. C. A. Newman, Class. Quantum Grav. {\bf 3} (1986), 527.
 \bibitem{Joshi-Dwivedi}
   P. S. Joshi and I. H. Dwivedi, Phys. Rev. {\bf D47} (1993), 5357.
 \bibitem{Jhingan:1997ia}
  S. Jhingan and P. S. Joshi, Annals of Israel Physical Society
  Vol 13. (1997), 357. 

 \bibitem{nsn1993}
   T.~Nakamura, M.~Shibata and K.~Nakao,
   Prog. Theor. Phys. {\bf 89} (1993), 821.

 \bibitem{Shapiro}
   S. L. Shapiro and S. A. Teukolsky, Phys. Rev. Lett. {\bf 66} (1991), 994;
   Phys. Rev.  {\bf D45} (1992), 2006.

 \bibitem{Echeverria:1993wf}
   F.~Echeverria,
   Phys. Rev. {\bf D47} (1993), 2271.

 \bibitem{Chiba}
   T.~Chiba,
   Prog. Theor. Phys. {\bf 95} (1996), 321.

 \bibitem{Iguchi} 
   H. Iguchi, K. Nakao, and T. Harada, Phys. Rev. {\bf D57} (1998), 7262.
 \bibitem{Penrose69}
   R. Penrose, Riv. Nuovo Cim. {\bf 1} (1969), 252.
 \bibitem{Penrose79}
   R. Penrose, in {\it General
   Relativity, an Einstein Centenary Survey}, ed. S. W. Hawking and
   W. Israel (Cambridge University Press, 1979), p. 581.
 \bibitem{Harada}
   T. Harada, H. Iguchi, and K. Nakao, Phys. Rev. {\bf D58} (1998), 041502.
 \bibitem{MTW}
   C. W. Misner, K. S. Thorne and J. A. Wheeler,
  {\it Gravitation} (Freeman, San Francisco, 1973).
 \bibitem{Misner}
   C. W. Misner and D. H. Sharp, Phys. Rev. {\bf 136} (1964), B571.
 \bibitem{Singh-Joshi}
   T. P. Singh and P. S. Joshi, Class. Quantum Grav. {\bf 13} (1996), 559.
 \bibitem{Jhingan}
   S. Jhingan, P. S. Joshi, and T. P. Singh, Class. Quantum Grav. {\bf
   13} (1996), 3057.
 \bibitem{Regge}
   T. Regge and J.A. Wheeler, Phys. Rev. {\bf 108} (1957), 1063. 
 \bibitem{Gerlach-Sengupta}
   U. H. Gerlach and U. K. Sengupta, Phys. Rev. {\bf D19} (1979), 2268.
 \bibitem{Chandra}
   S. Chandrasekhar and S. Detweiler, Proc. R. Soc. London {\bf 344} (1975),
   441.
 \bibitem{Landau} 
   L. D. Landau and E. M. Lifshitz, {\it The Classical Theory of Fields} 
   (Pergamon, London, 1975).
\end{thebibliography}
\end{document}